\documentstyle[sprocl,graphicx]{article}

\def\Journal#1#2#3#4{{#1} {\bf #2}, #3 (#4)}

\def\NPA{{\em Nucl. Phys.} A}

\def\PRC{{\em Phys. Rev.} C}

\def\A&A{{\em Astron. Astrophys.}}
\def\ApJ{{\em Astrophys. Journ.}}

\def\be{\begin{equation}}
\def\ee{\end{equation}}
\def\bea{\begin{eqnarray}}
\def\eea{\end{eqnarray}}

\begin{document}

\title{STRANGE HADRONIC STELLAR MATTER WITHIN THE BRUECKNER-BETHE-GOLDSTONE
THEORY}

\author{M. BALDO, {\underline {G. F. BURGIO}}}

\address{INFN Sezione di Catania, 57 Corso Italia, I-95129 Catania, Italy\\
and \\ ECT*, 286 Strada delle Tabarelle, I-38050 Trento, Italy 
\\E-mail: baldo@ct.infn.it, burgio@ct.infn.it} 

\author{H.-J. SCHULZE}

\address{Departament d'Estructura i Constituents de la Mat\'eria,
Universitat de Barcelona,\\ Av. Diagonal 647, E-08028 Barcelona,
Spain\\E-mail: schulze@ecm.ub.es}


\maketitle\abstracts{In the framework of the non-relativistic 
Brueckner-Bethe-Goldstone theory, we derive a microscopic equation of state 
for asymmetric and $\beta$-stable matter containing 
$\Sigma^-$ and $\Lambda$ hyperons. We mainly study the effects of three-body 
forces (TBF's) among nucleons on the hyperon formation and the 
equation of state (EoS). We find that, when TBF's are included, the stellar
core is almost equally populated by nucleons and hyperons. The resulting EoS,
which turns out to be extremely soft, has been used in order to calculate 
the static structure of neutron stars. We obtain a value of the maximum mass 
of 1.26 solar masses (1 solar mass $M_o \simeq 1.99 \cdot 10^{33} g$). 
Stellar rotations increase this value by about $12 \%$.}

\section{Neutron stars within the BBG approach}

The nuclear matter equation of state (EoS) is the fundamental input
for building models of neutron stars. These compact objects, among the
densest in the universe, are indeed characterized by values of the 
density which span from the iron density at the surface up to eight-ten 
times normal nuclear matter density in the core. Therefore a detailed 
knowledge of the equation of state over a wide range of densities 
is required \cite{shapiro}.
This is a very hard task from the theoretical point of view. 
In fact, whereas at densities close to the saturation value the matter 
consists mainly of nucleons and leptons, at higher densities several species 
of particles may appear due to the fast rise of the nucleon chemical 
potentials. 
In our work we perform microscopic calculations of the nuclear matter 
EoS containing fractions of $\Lambda$ and $\Sigma^-$ hyperons in the 
framework of the Brueckner-Hartree-Fock (BHF) scheme \cite{baldo}. 
The BHF approximation,
with the continuous choice for the single particle potential, reproduces 
closely the many-body calculations up to the three hole-line level.
In this approach, the basic input is the two-body interaction.
We chose the Paris and the Argonne $v_{18}$ potential for the 
nucleon-nucleon (NN) part, whereas the Nijmegen soft-core model has been 
adopted for the nucleon-hyperon (NY) potential. 
No hyperon-hyperon interaction is taken 
into account, since no robust experimental data are available yet.
For more details, the reader is referred to ref. \cite{bbs} and 
references therein.
However, as commonly known, all many-body methods fail to reproduce 
the empirical nuclear matter saturation point $\rho_0 = 0.17~fm^{-3}$. 
This drawback is commonly corrected 
by introducing three-body forces (TBF's) among nucleons. In our approach 
we have included a contribution containing a long range two-pion
exchange attractive part and an intermediate range repulsive part \cite{bbb}.
This allows the correct reproduction of the saturation point.
In figure 1 we show the chemical composition of $\beta$-stable 
and asymmetric nuclear matter containing hyperons (panel (a)) and  
the corresponding equation of state (panel (b)). The shown calculations 
have been performed using the Paris potential. We observe that hyperon
formation starts at densities $\rho \simeq 2-3$ times normal nuclear matter
density. The $\Sigma^-$ baryon appears earlier than
the $\Lambda$, in spite of its larger mass, because of the negative charge.
The appearance of strange particles has two main consequences, 
i) an almost equal percentage of nucleons and hyperons are present in the 
stellar core at high densities and ii) a strong
deleptonization of matter, since it is energetically convenient to maintain
charge neutrality through hyperon formation than $\beta$-decay.
The equation of state is displayed in panel (b). The dotted line represents
the case when only nucleons and leptons are present in stellar matter,
whereas the solid line shows the case when hyperons are included as well.
In the latter case the equation of state gets very soft,
since the kinetic energy of the already present baryonic species is converted
into masses of the new particles, thus lowering the total pressure.  
This fact has relevant consequences for the structure of the neutron stars. 

\begin{figure}
\includegraphics[totalheight=11cm,angle=270,bb=110 110 458 707]{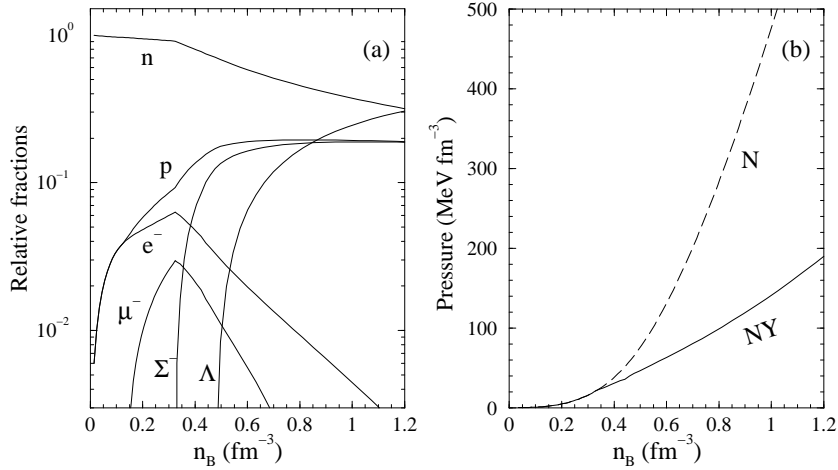}
\caption{In panel (a) we display the equilibrium composition of asymmetric and 
$\beta$-stable nuclear matter containing $\Sigma^-$ and $\Lambda$ hyperons. 
In panel (b) the solid(dotted) line represents the EoS obtained in the 
case when nucleons plus hyperons (only nucleons) are present.
\label{fig:bologna1}}
\end{figure}

\begin{figure}
\includegraphics[totalheight=11cm,angle=270,bb=145 112 494 676]{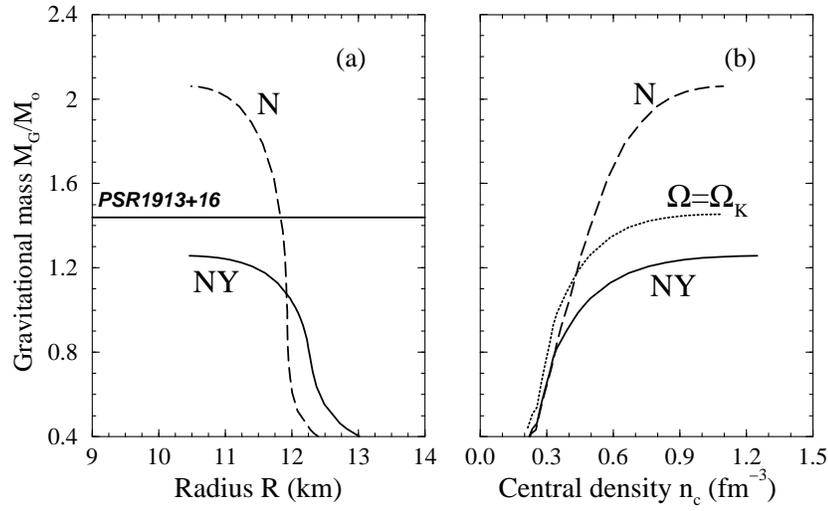}
\caption{In panel (a) the mass-radius relation is shown in the case of 
beta-stable matter with hyperons (solid line) and without hyperons 
(dashed line). The thick line represents the measured value of the pulsar 
PSR1913+16 mass. In panel (b) the mass is displayed
vs. the central density. The dotted line represents the equilibrium 
configurations of neutron stars containing nucleons plus hyperons and 
rotating at the Kepler frequency $\Omega_K$.\label{fig:fig2}}
\end{figure}

\section{Equilibrium configurations of neutron stars}

We assume that a star is a spherically symmetric distribution of mass in
hydrostatic equilibrium. The equilibrium configurations are obtained
by solving the Tolman-Oppenheimer-Volkoff (TOV) equations \cite{shapiro} for 
the pressure $P$ and the enclosed mass $m$,
\begin{eqnarray}  
  {dP(r)\over{dr}} &=& -{ G m(r) \rho(r) \over r^2 } \,
  {  \left[ 1 + \frac{P(r)}{\rho(r)} \right] 
  \left[ 1 + \frac{4\pi r^3 P(r)}{m(r)} \right] 
  \over
  1 - \frac{2 G m(r)}{r} } \:,
\\
  {dm(r) \over dr} &=& 4 \pi r^2 \rho(r) \:,
\end{eqnarray}
being $G$ the gravitational constant (we assume $c=1$). 
Starting with a central mass density $\rho(r=0) \equiv \rho_c$,  
we integrate out until the pressure on the surface equals the one 
corresponding to the density of iron.
This gives the stellar radius $R$ and the gravitational mass is then 
\begin{equation}
M_G~ \equiv ~ m(R)  = 4\pi \int_0^R dr\,r^2 \rho(r) \:. 
\end{equation}
For the outer part of the neutron star we have used the equations of state
by Feynman-Metropolis-Teller \cite{fey} and Baym-Pethick-Sutherland 
\cite{baym}, and for the medium-density regime  
we use the results of Negele and Vautherin \cite{nv}. 
For density $\rho > 0.08\,{\rm fm}^{-3}$ 
we use the microscopic equations of state obtained in the BHF approximation
described above. For comparison, we also perform calculations of neutron 
star structure for the case of asymmetric and $\beta$-stable nucleonic matter.
The results are plotted in Fig.2. We display the gravitational mass $M_G$ 
(in units of the solar mass $M_o$)
as a function of the radius $R$ (panel (a)) and central baryon density 
$n_c$ (panel (b)). We note that the inclusion of hyperons lowers the 
value of the maximum mass from about 2.1 $M_o$ down to 1.26 $M_o$.
This value lies below the value of the best 
observed pulsar mass, PSR1916+13, which amounts to 1.44 solar masses.
However the observational data can be fitted if rotations are included, see
dotted line in panel (b). In this case only equilibrium configurations 
rotating at the Kepler frequency $\Omega_K$ are shown.\par\noindent
In conclusion, the main finding of our work is the surprisingly low value
of the maximum mass of a neutron star, which hardly comprises the 
observational data. This fact indicates how sensitive the properties of the
neutron stars are to the details of the interaction. In particular
our result calls for the need of including realistic hyperon-hyperon 
interactions.


\section*{References}
\begin{thebibliography}{99}
\bibitem{shapiro}
   S. L. Shapiro and S. A. Teukolsky, 
   {\em Black Holes, White Dwarfs and Neutron Stars}  
   (John Wiley \& Sons, New York, 1983)
\bibitem{baldo}
   M. Baldo, 
   {\em Nuclear Methods and the Nuclear Equation of State}
   (World Scientific, Singapore, 1999)  
\bibitem{bbs}
   M. Baldo, G. F. Burgio, and H.-J. Schulze,
   \Journal{\PRC}{61}{055801-1}{2000}. 
\bibitem{bbb}
   M. Baldo, I. Bombaci, and G. F. Burgio,
   \Journal {\A&A}{328}{274}{1997}.
\bibitem{fey}
   R. Feynman, F. Metropolis, and E. Teller,
   \Journal{\PRC}{75}{1561}{1949};
\bibitem{baym}
   G. Baym, C. Pethick, and D. Sutherland,
   \Journal{\ApJ}{170}{299} {1971}.
\bibitem{nv}
   J. W. Negele and D. Vautherin,
   \Journal{\NPA}{207}{298}{1973}.

\end{thebibliography}
\end{document}